\documentclass[twocolumn,showpacs,amssymb,aps,prl]{revtex4}

\usepackage{graphicx}
\usepackage{bm}

\begin{document}
\title{Observation of Apparently Zero-Conductance States in Corbino Samples}
\author{C. L. Yang}
\author{M. A. Zudov}
\author{T. A. Knuuttila}
\author{R. R. Du}
\affiliation{Department of Physics, University of Utah, Salt Lake City, Utah 84112}
\author{L. N. Pfeiffer}
\author{K. W. West}
\affiliation{Bell Laboratories, Lucent Technologies, Murray Hill, New Jersey 07974}
\date{March 21, 2003}
\begin{abstract}
Using Corbino samples we have observed oscillatory conductance in a high-mobility two-dimensional electron system subjected to crossed microwave and magnetic fields. On the
strongest of the oscillation minima the conductance is found to be
vanishingly small, possibly indicating an insulating state associated
with these minima.
\end{abstract}
\maketitle

Following the discovery by Zudov et al \cite{zudovprb} of giant amplitude
magnetoresistance oscillations in a two-dimensional electron system (2DES)
subjected to crossed microwave (MW) and weak magnetic ($B$) fields, observations
of exponentially small resistance states (ESRS) associated with the oscillation minima
were reported \cite {mani, zudovzrs}. The emergence of 
ESRS at the oscillation minima strongly correlates with an increasing electron mobility in the 2DES \cite{firstzrs}.  The origin 
of the oscillations and the mechanism leading to ESRS are of considerable current theoretical
interest \cite{phillips, durst, andreev, anderson, shi, koulakov, volkov, mikhailov}.
Several models based on oscillatory density of states and negative d.c.~response of
the 2DES to a driving a.c.~field are proposed \cite{durst,andreev,anderson,shi}, possibly providing a
simple and physically transparent mechanism for the oscillations. The ESRS, it follows,
are thought to result from an instability caused by the negative response and subsequent
redistribution of electric currents \cite{andreev,anderson}. Other models consider
either the electron orbital dynamics and the formation of a sliding charge-density
wave \cite{phillips}, or an interplay between the bulk and edge magnetoplasmon
modes which could lead to oscillations \cite{mikhailov}. A classical mechanism based
on non-parabolicity of the electron dispersion is proposed in Ref. \cite{koulakov}.

\begin{figure}[!hb]
\includegraphics{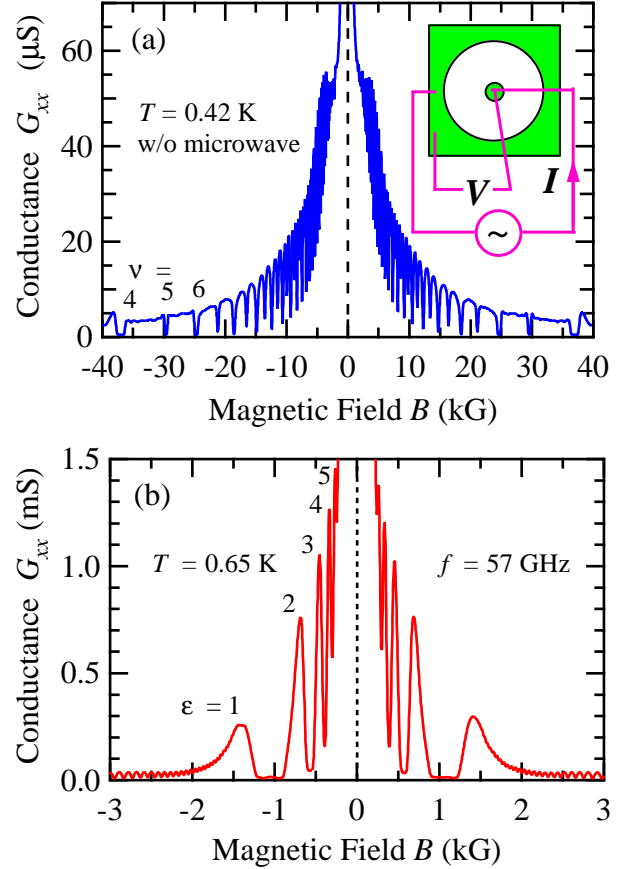}
\caption{
(a) Magnetoconductance of a Corbino sample (without MW illumination) is shown to exhibit
sharp SdH oscillations at low magnetic field and vanishing IQHE minima at high magnetic
field. The insert depicts the geometry of the sample and the measurement circuit.
(b) The conductance oscillations observed in the Corbino sample with microwave illumination.
A vanishing-conductance state at the first minimum is observed.}
\label{fig1}
\end{figure}

While the electronic d.c.~transport in this regime is controlled by an
a.c. MW field, the diagonal resistance measured in ESRS resembles remarkably
that observed in the quantized Hall effect (QHE), a purely d.c.~transport phenomenon
pertaining to a strong magnetic field. It is well known \cite{tsuiprb82} that,
in the QHE a vanishing resistivity, $\rho_{xx}$, is equivalent to a vanishing
conductivity, $\sigma_{xx}$, and these two quantities relate to each other by
\begin{equation}
\sigma_{xx}=\frac{\rho_{xx}}{\rho_{xx}^2+\rho_{xy}^2}
\approx \rho_{xx}/\rho_{xy}^2= (n_ee/B)^2\rho_{xx},
\label{tensor}
\end{equation}
where $n_e$ is the electron density. The vanishing conductivity in the QHE
is taken as the evidence for an insulating state, and the rigidity of the state 
can be understood by the formation of an energy gap, either of single-electron
origin (integer QHE) or of many-electron origin (fractional QHE).

We report on MW-induced vanishing d.c.~\emph{conductance} observed in Corbino 
samples of a high-mobility 2DES. The experiments demonstrate that, regardless of 
the presence of the MW fields, the conductance and resistance are invertible according to
Eq.~(1), up to a scaling factor; hence, the observed vanishing-conductance states (VCS) 
correspond to ESRS. Such observation from a topologically distinct sample 
could impose new constraint on theoretical models of ESRS.

Our samples were cleaved from a Al$_{0.24}$Ga$_{0.76}$As/ GaAs/Al$_{0.24}$Ga$_{0.76}$As
quantum well (QW) wafer grown by molecular beam epitaxy. The width of the QW is 25 nm
and the electrons are provided by Si $\delta$-doping layers 80 nm above and below the QW.
After illumination by a red light-emitting diode at $T\approx 1.5$ K, the electron density,
 $n_e$, and mobility, $\mu$, reached $3.55\times 10^{11}$ cm$^{-2}$ and
$12.8\times10^6$ cm$^2$/Vs, respectively. The Corbino samples, with an inner diameter
$r_1\approx 0.5$ mm and an outer diameter $r_2\approx 3.0$ mm, were made on
a  $\sim$4 mm$\times$4 mm square. Ohmic contacts were made of indium. To compare the 
conductance measurement with a resistance measurement,
a $\sim$4 mm$\times$4 mm square sample was made with eight indium contacts
along the perimeter. The measurements were performed in a sorption-pumped
 $^3$He cryostat equipped with a superconducting magnet; the magnetic field was
calibrated using a low temperature Hall probe. The microwaves were generated by Gunn diodes
and guided down to the sample via an oversized (WR-28) waveguide.
The conductance or resistance traces were recorded employing a low-frequency
(2.7 Hz) lock-in technique while the sample was immersed in $^3$He liquid and
under continuous microwave illumination of fixed frequency, $f$, and power, $P$.

The diagonal conductance, $G_{xx} = I/V$, of the Corbino sample was obtained by
measuring the current, $I$, passing through the 2DES, while applying a 
voltage, $V$, between the inner and outer contacts.
Here $x$ denotes the direction along the radius. A typical bias of $V \sim 1$ mV
was used for the measurements. The inset of Fig.~1(a) shows a schematic circuit of 
the measurement.

Without the MWs and while sweeping the magnetic field, the $G_{xx}$ trace shows sequentially,
in Fig.~1(a),  a Drude conductance around $B = 0$, sharp Shubnikov--de Haas (SdH)
oscillations at $B \gtrsim 1.5$ kG, and vanishing conductance at the integer 
QHE (IQHE) minima at $B \gtrsim 10$ kG. The trace is strictly symmetrical with
respect to $B = 0$, indicating that the recorded $G_{xx}$ is free of any mixture
with the Hall conductance. Altogether, such standard d.c.~magnetotransport
data attests exceptional quality of the Corbino sample.
We note that at this temperature a residual conductance in the IQHE remains
measurable. Its value is typically $\lesssim 5 \times 10^{-7}$ S, which is $\sim$$10^{-7}$
of the conductance at $B = 0$ ($G_{xx}(0)\sim 2.5$~S, converted from mobility).
Empirically we take this value as the lower limit for determining the MW-induced
vanishing conductance, presented below.

Fig.~1(b) shows a $G_{xx}$ trace with MW illumination of $f = 57$ GHz and with an incident
power $P\approx10\ \mu$W on the sample surface. Notice that the temperatures
marked in both (a) and (b) are those measured in the $^3$He liquid.
Strong MW-induced conductance oscillations up to 5 orders are observed. The peaks are marked
by $\epsilon \equiv \omega/\omega_c =1,\ 2,\ 3,\ \ldots$, where $\omega_c =eB/m^*$ is the
cyclotron frequency, $m^*=0.068\,m_e$ is the effective mass of the electron,
and $\omega =2\pi f$. 

Our central finding from such measurements, however, is the 
vanishing-conductance state observed at the strongest oscillation minimum, around $B =\pm 1.05$ kG. 
Such state spans a wide range of Landau level filling factor, $\nu =n_{e}h/eB$, 
from $\nu\sim160$ to $\nu\sim120$. At $T=0.65$ K, a residual conductance,   
$G_{xx}\lesssim 2 \times10^{-5}$ S, was detected in the VCS regime, which is
$\sim$$10^{-5}$ of the conductance at $B=0$. Considering the geometric factor of the Corbino sample
and assuming an uniform distribution of the electric current passing through the contacts, such
conductance value corresponds to a residual conductivity $\ll e^2/h$. Finite residual conductance
has also been observed in the IQHE regime, which can be attributed to thermally activated 
conduction \cite{sds_1993} or variable-range hopping conduction \cite{hopping}. At this stage we 
are not able to rule out the possibility that the residual conductance of the VCS is caused by MW-induced 
parallel conduction. While the origin of the residual conductance remains to be clarified, it appears 
that the 2DES behaves like an insulator in the VCS regime.

\begin{figure}
\includegraphics{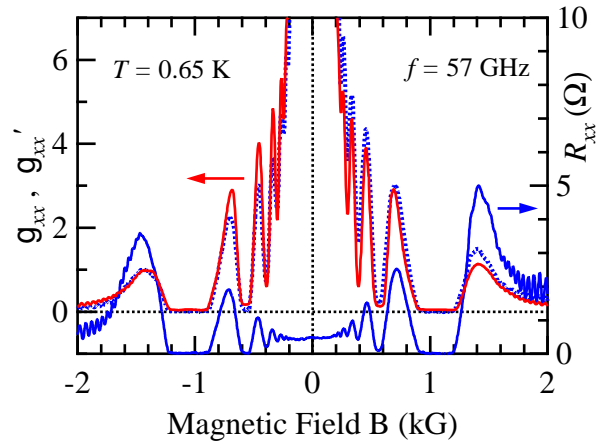}
\caption{\label{fig2}
The measured $g_{xx}$ (normalized conductance, as defined in the text) of a Corbino sample 
is plotted together with a  $g_{xx}'$ (dotted line) inverted via Eq. (1) from $R_{xx}$ measured
on a square sample. An excellent agreement between the measured $g_{xx}$ and the inverted
one is found, indicating that the microwave-induced resistance oscillations and
the conductance oscillations are invertible according to standard d.c.~transport
tensor relation.}
\end{figure}

In order to compare the MW-induced conductance oscillations with the resistance oscillations,
we calculate a diagonal conductance, $G_{xx}'$, from a diagonal resistance, $R_{xx}$,
measured on a square sample, using the inversion relation
$G_{xx}'\approx R_{xx}/\rho_{xy}^2= (n_ee/B)^2R_{xx}$.
The parameter which can be directly compared between different samples is not the conductance
but the conductivity. Since the conductivity is proportional to conductance with a factor depending
on the sample dimensions, which are not accurately known, we can normalize the conductance to a
specific point $B_0$ (e.g., the first maximum of the oscillations), and compare the normalized 
conductances $g_{xx}$ ($\equiv G_{xx}(B)/G_{xx}(B_0)$). Fig. 2 displays
both the $g_{xx}$ measured from the Corbino sample and the $g_{xx}'$ converted from $R_{xx}$
measured on a square sample. The $R_{xx}$ of the square sample is also shown in
Fig. 2, which was measured under the same conditions as the 
$G_{xx}$ ($T = 0.65$ K, $f  = 57$ GHz, $P\approx 10$ $\mu$W). Slight
asymmetry of the $R_{xx}$ with respect to $B=0$ might indicate a weak mixture of
resistance tensor elements. Excellent agreement between the $g_{xx}$'s clearly demonstrates
that, under MW illumination the d.c.~conductance and resistance remain invertible up 
to a scaling factor.

\begin{figure}
\includegraphics{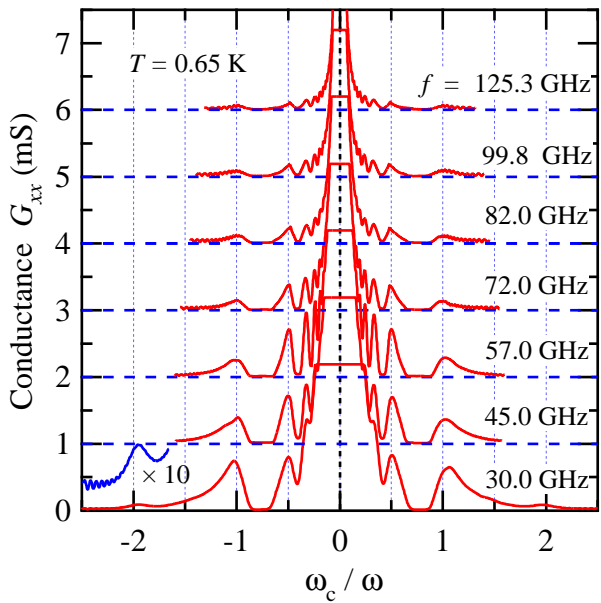}
\caption{\label{fig3}
The conductance oscillations of the Corbino sample for selected MW frequencies
with roughly the same $P\approx 10\ \mu$W and $T\approx 0.65\ $K (for clarity,
traces are vertically shifted in steps of 1 mS).
$m^*=0.068\,m_e$ is used to scale the magnetic field.
For all frequencies, strong oscillations and the vanishing-conductance state
are observed. The strength of the oscillation is decreasing with increasing MW frequency,
which could be due to the decreasing of the photon number.
Additional maximum at $\omega_c/\omega \approx 2$ is observed for $f < 40$ GHz,
as shown by the enlarged trace at negative $B$ side for $f = 30$ GHz.}
\end{figure}

In addition, we have measured the conductance at different MW frequencies
(from 25.5 GHz  to 130 GHz) but at roughly the same MW power $P\approx10\ \mu$W
(at the sample surface) and the same temperature $T\approx 0.65$ K. Selected conductance
traces against the scaled magnetic field, $1/\epsilon = \omega_c /\omega$, are
shown in Fig. 3. Strong oscillations as well as  the VCS are
observed at all MW frequencies. Within experimental accuracy, no phase shift
is observed for the major peaks ($\epsilon = 1,\ 2,\ 3$). We also
notice a trend in which the strength of the
oscillations, as measured by the peak height, is decreasing with increasing frequency.
Such an observation can be qualitatively accounted for by the number of photons incident
on the 2DES. Since the MW power is roughly the same,
the number of photons is inversely proportional to the MW frequency, leading to a diminishing
of the oscillations at higher $f$. For $f < 40$ GHz, an additional maximum
at $1/\epsilon \approx 2$ is observed; an example can be seen on both positive and negative 
$B$ of the $f = 30$ GHz trace. Such additional peaks have been previously seen 
in the ESRS experiments \cite{zudovzrs}.

In conclusion, we have observed MW-induced vanishing-conductance states in a high-mobility
2DES of Corbino geometry. While the effect is driven by an a.c.~field, the d.c.~conductivity 
and the resistivity are found to be invertible using the standard d.c.~transport tensor 
relation. The observation of the vanishing-conductance states opens 
an unique experimental window for further investigations, and could impose new
constraint on theoretical models of ESRS.

We thank H. L. Stormer, R. L. Willett,  H. W. Jiang, and S. Hill for valuable discussions
on experiments. The work at the University of Utah was supported by a DARPA QUIST grant.
T. A. K. and R. R. D. were also supported in part by DOE.

\bibliographystyle{unsrt}

\end{document}